\documentclass[preprint,12pt]{elsarticle}

\makeatletter
\def\ps@pprintTitle{%
 \let\@oddhead\@empty
 \let\@evenhead\@empty
 \def\@oddfoot{}%
 \let\@evenfoot\@oddfoot}
\makeatother

\usepackage{amssymb}

\usepackage{lineno}

\usepackage{mwe}
\usepackage{subfig}
\usepackage{float}
\usepackage{indentfirst}
\usepackage[export]{adjustbox}
\usepackage{tabu}
\newcolumntype{C}[1]{>{\centering\let\newline\\\arraybackslash\hspace{0pt}}m{#1}}
\usepackage{graphicx}

\usepackage{hyperref}
\usepackage{textpos}

\begin{document}

\begin{frontmatter}



\title{	Improvement of heat exchanger efficiency by using hydraulic and thermal entrance regions}


\author[add1]{Alexey Andrianov}
\author[add1]{Alexander Ustinov}
\author[add1,add2]{Dmitry Loginov}

\address[add1]{Skolkovo Institute of Science and Technology, Moscow, Russia}
\address[add2]{Moscow Institute of Science and Technology, Dolgoprudny, Russia}

\begin{abstract}
This study investigates one of the possible approaches of improvement of heat exchangers efficiency. This problem was and still is a burning issue due to a wide variety of applications: aviation and aerospace, energy sector, chemical and food industry, refrigeration and cryogenic engineering, heating/cooling and conditioning services, thermal engines, housing and utility infrastructure etc. Among different types of the heat exchangers, shell-and-tube unit was chosen for the research, as its part in the overall number of heat exchangers is about 90\% [3]. \par
Literature review shows that most approaches of improvement are based on the heat transfer surface increasing and laminar-to-turbulent flow transition using different types of riffles forming and shaped inserts. In this article, a novel approach to the heat transfer intensification was employed. The main hypothesis is that applying of multi-chamber design of heat exchanger -- ordinary shell-and-tube regions intersperse with common for all tubes regions -- will help to improve the utilization of the entrance hydraulic and thermal regions thereby receive higher heat transfer coefficients and higher heat capacity of the heat exchange device. To prove the hypotheses we take the following steps. Firstly, development of the new geometry of the heat-exchanger design - multi chambers construction. Secondly, proving of the higher efficiency of novel design comparing to ordinary design by analytical calculations. Thirdly, numerical simulation of the heat exchange process and fluids flow in both types of heat exchangers that proves the analytical solution. \par


\end{abstract}

\begin{keyword}
Energy systems \sep Heat exchanger \sep CFD modelling

\begin{textblock*}{200mm}(0\textwidth,1cm)
\underline{Email addresses}: \\ aleksei.andrianov@skolkovotech.ru (Aleksei Andrianov),\\ alexander.ustinov@gmail.com (Alexander Ustinov), \\ dmitry.loginov@skolkovotech.ru (Dmitry Loginov).
\end{textblock*}


\end{keyword}

\end{frontmatter}


\section{Introduction}
\label{S:1}

Technical-and-economic indexes of the different types of heating plants depend heavily on the heat exchangers [1, 2, 4, 9, 11]. The reason is significant part of the heat exchange units in the overall mass of the system. Furthermore, industrial growth in terms of both volume and capacity leads to the increase of heat exchangers capacity, overall dimensions and weight. As a result, the cost of the whole system increases. That is why the problem of rationalization and further development of the heat exchangers becomes more and more important and challenging issue.\par
This challenge is important for a wide range of the industries due to the fact, that there are many different types of heat exchange units and they are essential part in such areas as: aviation and aerospace, energy sector, chemical and food industry, refrigeration and cryogenic engineering, heating/cooling and conditioning services, thermal engines, housing and utility infrastructure etc [6, 7].\par
Many researchers paid attention to the problem of the heat exchangers efficiency improvement [5], describing the promising ways of the heat transfer intensification [8, ]. There are two main principles of the heat capacity increase. First, to increase the heat transfer surface area, allowing thermal energy change from the greater surface. In this case, a great variety of fins of different shapes and locations provides the better heat performance of the system. Second, to make the turbulence of the fluid flows. The laminar-to-turbulent transition can be achieved by using many kinds of shaped inserts in the inner space of the tubes, intensifying baffles of special design in the shell side -- space between tubes inside the shell -- of the heat exchanger, or special geometry of the tube itself for turbulent flow creation [12]. \par
The aim of the current study is the development of the novel approach in the intensification process of the heat transfer. 
New construction -- multi-chamber design -- will help to achieve greater heat transfer coefficient and thereby will allow getting greater heat capacity of the system. The reason for that is special geometry that provides better utilization of the entrance hydraulic and thermal regions.\par
This hypothesis is going to be proved by three interrelated steps. The first tool is analytical calculation of the heat capacity of the heat exchanger. The classical approach [10] is implemented for the ordinary type of the shell-and-tube heat exchanger. The novel unit capacity is estimated by dividing the system into sections and summing up the capacity of each section. The second tool is numerical solution of both types of the heat exchanger with the same boundary conditions. The result is the pressure and temperature drop of fluid flows in two systems. The third tool is experimental solution, using specially constructed laboratory unit for testing ordinary and novel heat exchangers.\par
Comparison of the results of all three steps for two systems will allow getting answers if the novel geometry has better heat performance. If so, how much the capacity of the new system is higher.

\section{Calculations of the ordinary design and the novel multi-chamber design heat exchangers}

In this section shell and tube heat exchangers in novel and ordinary designs will be proposed with different boundary conditions, liquid-liquid type. \par

The main hypothesis is that applying of multi-chamber design of heat exchanger -- ordinary shell-and-tube regions intersperse with common for all tubes regions -- will help to improve the utilization of the entrance hydraulic and thermal regions thereby receive higher heat transfer coefficients and higher heat capacity of the heat exchange device. \par

Ordinary design of the heat transfer unit is usual shell-and-tube heat exchanger, which consists of shell with outlet and inlet of the “cold” heat transfer agent. Tubes are inside the shell with the common space for all tubes on both sides.  These sides are inlet and outlet of the “hot” heat transfer agent. Values of all geometrical and thermal parameters are present below (Figure 1). Materials are water for hot and cold fluids; all walls are made of stainless steel. \\

\begin{figure}[H]
\centering
\includegraphics[width=0.65\textwidth]{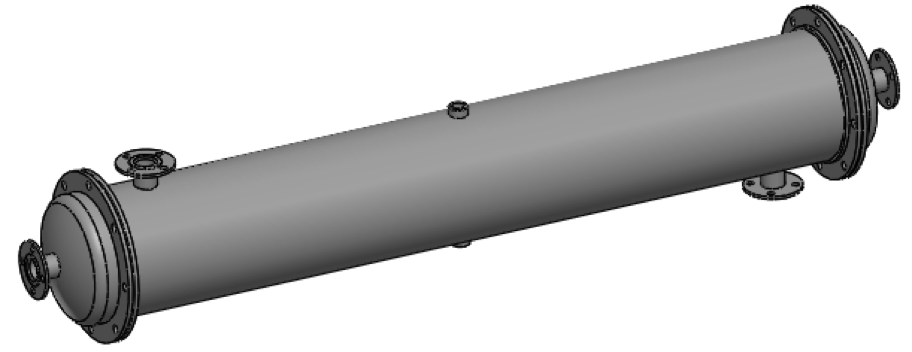}
\hfill
\includegraphics[width=0.65\textwidth]{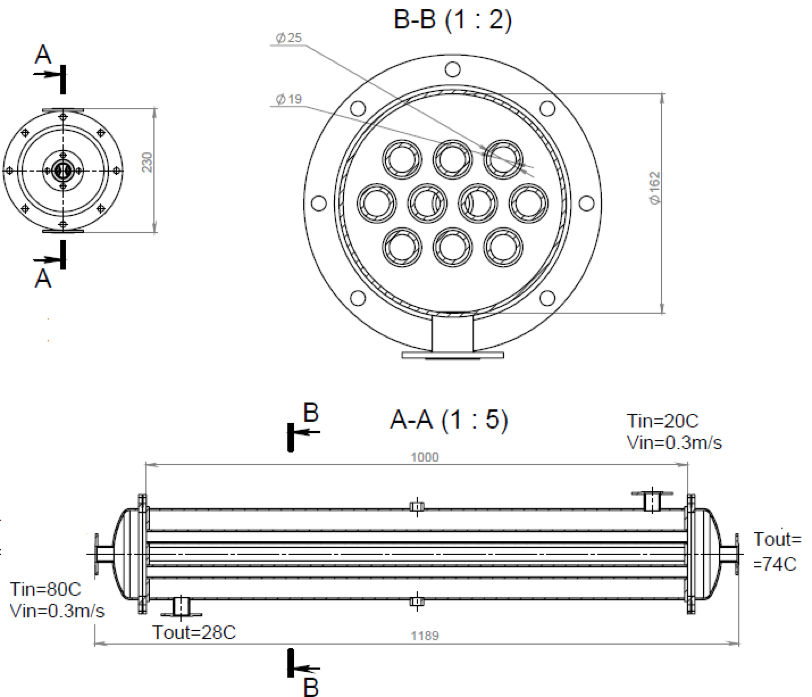}

\centering
\caption{Ordinary design of the shell-and-tube heat exchanger.}
\label{fig:dummy}

\end{figure}

Novel multi-chamber design is the development of the usual shell-and-tube heat exchanger. It has several qualitative changes. New heat exchange unit consist of three sections. Each section is the combination of two regions: first region is ordinary region of tubes inside the shell; second region is the common space for all tubes (Figure 2). This geometry can be described as the series of the three individual small shell-and-tube heat exchangers. This special design provides better utilization of the entrance hydraulic and thermal regions. The reason for that is relatively small length of each section -- fluid enters the pipes and boundary layers start to fill the inner space of the pipes until the end of the entrance region. Approximately at this point the common space for all tubes occurs. It means that we look at the process only in the entering regions where the highest heat transfer coefficient is. Values of all geometrical and thermal parameters are present below (Figure 2). \\

 \begin{figure}[H]
\centering
\includegraphics[width=0.7\textwidth]{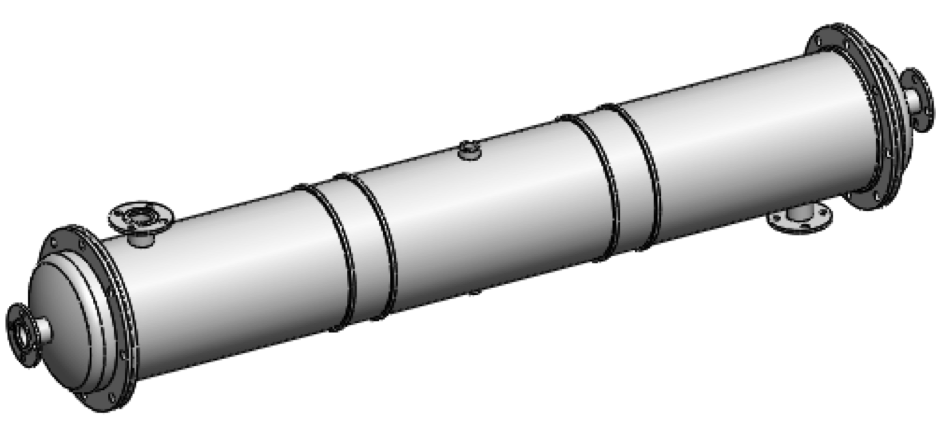}
\hfill
\includegraphics[width=0.7\textwidth]{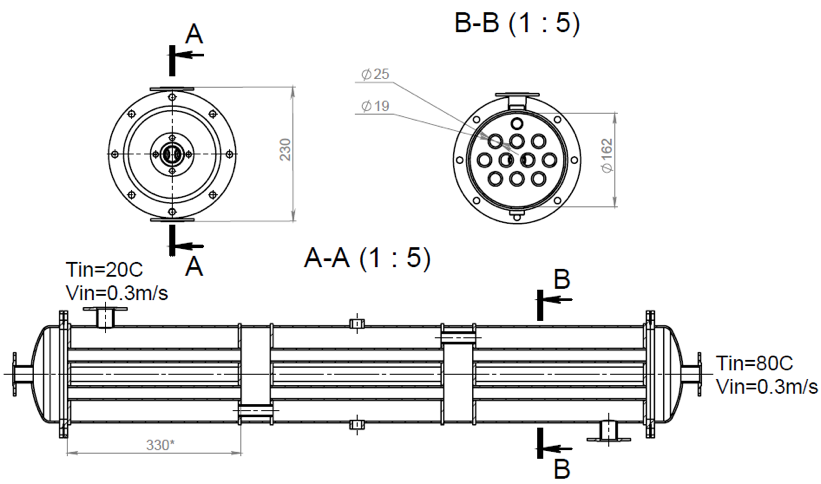}

\centering
\caption{Novel multi-chamber design of the shell-and-tube heat exchanger.}
\label{fig:dummy}

\end{figure}
\medskip

\subsection{Calculation of the heat capacity of the ordinary shell-and-tube heat exchanger}

We need to define the heat capacity of the system. It depends on area of the heat transfer surface, temperature difference and coefficient of the heat transmission. \medskip
The area of the heat transfer surface is sum of the outer surfaces of n number tubes:
$$F=n \cdot \pi \cdot d \cdot L=10 \cdot \pi \cdot 0.025 \cdot 1=0.785\ m^2$$

The logarithmic mean temperature difference is defined using equation: 
$$ \Delta T = \frac{\left( T^\prime_h - T^\prime_c \right) - \left( T^{\prime \prime}_h - T^{\prime \prime} _c \right) }{ln \frac{ \left( T^\prime _h-T^\prime _c \right) } { \left( T^{\prime \prime} _h - T^{\prime \prime} _c \right) }}  = \frac{ \left( 80 - 20 \right) - \left( 74 - 28 \right) }{ln \frac{ \left( 80 - 20 \right) }{ \left( 74 - 28 \right) }} = 52.7\ ^{\circ}\mathrm{C}  $$ 

We need to make preliminary calculations to define coefficient of the heat transmission: \par

First, define the flow regime, using Reynolds number, for the inner tubes space and shell side. For the tubes, the characteristic dimension is equal to the inner diameter of the tube. For the shell side, the characteristic dimension is defined using hydraulic diameter equation. 
$$l_{tube}=d_{in}=0.019 m$$
$$l_{shell}=d_{eq}=\frac{4 \cdot f}{ \Pi} = \frac{4 \cdot (\pi \cdot (D_{in}/2)^2-n \cdot \pi \cdot (d/2)^2 )}{\pi \cdot D_{in}+ \pi  \cdot d  \cdot n}=0.0465\ m$$  


Reynolds number for the inner tube space -- hot heat transfer agent:
$$Re_{tube}=\frac{w_{tube} \cdot l_{tube}} {v_{tube}} =\frac{0.3 \cdot 0.019}{ 3.62 \cdot 10^{-7} }=15707$$

Reynolds number for the shell side -- cold heat transfer agent:
$$Re_{shell}=\frac{w_{shell} \cdot l_{shell}}{v_{shell}} =\frac{0.3 \cdot 0.0465}{9.19 \cdot 10^{-7} }=15183$$


The values of the Reynolds number both for inner tube space and shell side are higher than 10 000. It means that the flow regime is turbulent and we need equation for Nusselt number [13, 15, ] in the turbulent flow. 

Prandtl number for both hot and cold fluids in inner tube space and shell side has the following form: 
$$Pr_{tube}=\frac{c_{tube}  \cdot v_{tube}  \cdot \rho_{tube}}{\lambda_{tube}} =\frac{4194 \cdot 3.62 \cdot 10^{-7} \cdot 975}{0.68}=2.2$$
$$Pr_{shell}=\frac{c_{shell}  \cdot v_{shell}  \cdot \rho_{shell}}{\lambda_{shell}} =\frac{4175 \cdot 9.19 \cdot 10^{-7} \cdot 1000}{0.59}=6.4$$

Nusselt number for the turbulent flow has the following form in the inner tube space -- hot fluid:
$$Nu_{tube}=0.021 \cdot \varepsilon_1 \cdot Re^{0.8}  \cdot Pr^{0.43}  \cdot (Pr/Pr_w )^{0.25}=62.75 $$ 

Nusselt number for the turbulent flow has the following form in the shell side -- cold fluid:
$$Nu_{shell}=0.021 \cdot \varepsilon_1 \cdot Re^{0.8} \cdot Pr^{0.43}  \cdot (Pr/Pr_w )^{0.25}=106.2$$ 

After Nusselt number calculation it is possible to define the heat transfer coefficient both for “cold” and “hot” heat transfer agents from the equation for Nusselt number:

For the hot fluid in the inner tube space:
$$\alpha_1=\frac{Nu_{tube} \cdot \lambda_{tube}}{l_{tube}} =\frac{62.75 \cdot 0.68}{0.019} = 2248 \ W/(m^2 \cdot ^{\circ}\mathrm{C})$$

For the cold fluid in the shell side:
$$\alpha_2=\frac{Nu_{shell} \cdot \lambda_{shell}}{l_{shell}} =\frac{106.2 \cdot 0.59}{0.0465} = 1353 \ W/(m^2 \cdot ^{\circ}\mathrm{C})$$

Now we can find the value of the coefficient of the heat transmission:
$$k=\frac{1}{\frac{1}{\alpha_1} +\frac{\delta}{\lambda}+\frac{1}{\alpha_2}}=\frac{1}{\frac{1}{2248}+\frac{0.003}{20}+\frac{1}{1353}} = 749 \ W/(m^2 \cdot ^{\circ}\mathrm{C})$$

Finally, we can define the heat capacity of the heat exchanger:
$$Q = 749 \cdot 52.7 \cdot 0.785=31029 \ W$$ \medskip

\subsection{Estimation of the heat capacity of the novel multi-chamber shell-and-tube heat exchanger}

This paragraph called “estimation” rather than “calculation” due to specific cause. As the design of the multi-chamber heat exchanger is new, there is no special technique to calculate its heat capacity like in the previous paragraph. Therefore, we can only estimate it taken into consideration several assumptions. \par

First, we will consider each section of the novel heat exchanger as a separate small ordinary shell-and-tube heat exchanger. Second, the distribution of the temperature between inlet region and outlet region will be taken as uniform -- the same temperature drop in each section. \par

The temperature drop in the previous paragraph is  $6^{\circ}\mathrm{C}$ for hot fluid domain and  $8^{\circ}\mathrm{C}$ for cold fluid domain. For novel heat exchanger estimation, we will take the same temperature drops, paying attention to the distribution mentioned above. Therefore, the temperature drop in each section for the hot heat transfer agent will be  $2^{\circ}\mathrm{C}$; for the cold heat transfer agent -- $2.6^{\circ}\mathrm{C}$.

The area of the heat transfer surface is sum of the outer surfaces of n number tubes:
$$F=n \cdot \pi \cdot d \cdot L=10 \cdot \pi \cdot 0.025 \cdot 1=0.2615\ m^2$$

The logarithmic mean temperature difference is defined using equation: 
$$ \Delta T = \frac{\left( T^\prime_h - T^\prime_c \right) - \left( T^{\prime \prime}_h - T^{\prime \prime} _c \right) }{ln \frac{ \left( T^\prime _h-T^\prime _c \right) } { \left( T^{\prime \prime} _h - T^{\prime \prime} _c \right) }}  = \frac{ \left( 80 - 20 \right) - \left( 74 - 28 \right) }{ln \frac{ \left( 80 - 20 \right) }{ \left( 74 - 28 \right) }} = 57.66\ ^{\circ}\mathrm{C}  $$ 

We need to make preliminary calculations to define coefficient of the heat transmission: \par

First, define the flow regime, using Reynolds number, for the inner tubes space and shell side. For the tubes, the characteristic dimension is equal to the inner diameter of the tube. For the shell side, the characteristic dimension is defined using hydraulic diameter equation. 
$$l_{tube}=d_{in}=0.019 m$$
$$l_{shell}=d_{eq}=\frac{4 \cdot f}{ \Pi} = \frac{4 \cdot (\pi \cdot (D_{in}/2)^2-n \cdot \pi \cdot (d/2)^2 )}{\pi \cdot D_{in}+ \pi  \cdot d  \cdot n}=0.0465\ m$$ 

Reynolds number for the inner tube space -- hot heat transfer agent:
$$Re_{tube}=\frac{w_{tube} \cdot l_{tube}} {v_{tube}} =\frac{0.3 \cdot 0.019}{ 3.62 \cdot 10^{-7} }=16144$$

Reynolds number for the shell side -- cold heat transfer agent:
$$Re_{shell}=\frac{w_{shell} \cdot l_{shell}}{v_{shell}} =\frac{0.3 \cdot 0.0465}{9.19 \cdot 10^{-7} }=14257$$


The values of the Reynolds number both for inner tube space and shell side are higher than 10 000. It means that the flow regime is turbulent and we need equation for Nusselt number in the turbulent flow. 

Prandtl number for both hot and cold fluids in inner tube space and shell side has the following form: 
$$Pr_{tube}=\frac{c_{tube}  \cdot v_{tube}  \cdot \rho_{tube}}{\lambda_{tube}} =\frac{4194 \cdot 3.62 \cdot 10^{-7} \cdot 975}{0.68}=2.16$$
$$Pr_{shell}=\frac{c_{shell}  \cdot v_{shell}  \cdot \rho_{shell}}{\lambda_{shell}} =\frac{4175 \cdot 9.19 \cdot 10^{-7} \cdot 1000}{0.59}=6.97$$

Nusselt number for the turbulent flow has the following form in the inner tube space -- hot fluid:
$$Nu_{tube}=0.021 \cdot \varepsilon_1 \cdot Re^{0.8}  \cdot Pr^{0.43}  \cdot (Pr/Pr_w )^{0.25}=63.26 $$ 

Nusselt number for the turbulent flow has the following form in the shell side -- cold fluid:
$$Nu_{shell}=0.021 \cdot \varepsilon_1 \cdot Re^{0.8} \cdot Pr^{0.43}  \cdot (Pr/Pr_w )^{0.25}=104.12$$ 

After Nusselt number calculation it is possible to define the heat transfer coefficient both for “cold” and “hot” heat transfer agents from the equation for Nusselt number:

For the hot fluid in the inner tube space:
$$\alpha_1=\frac{Nu_{tube} \cdot \lambda_{tube}}{l_{tube}} =\frac{62.75 \cdot 0.68}{0.019} = 2277  \ W/(m^2 \cdot ^{\circ}\mathrm{C})$$

For the cold fluid in the shell side:
$$\alpha_2=\frac{Nu_{shell} \cdot \lambda_{shell}}{l_{shell}} =\frac{106.2 \cdot 0.59}{0.0465} = 1316  \ W/(m^2 \cdot ^{\circ}\mathrm{C})$$

Now we can find the value of the coefficient of the heat transmission:
$$k=\frac{1}{\frac{1}{\alpha_1} +\frac{\delta}{\lambda}+\frac{1}{\alpha_2}}=\frac{1}{\frac{1}{2248}+\frac{0.003}{20}+\frac{1}{1353}} = 741.5 \ W/(m^2 \cdot ^{\circ}\mathrm{C})$$

Finally, we can define the heat capacity of the heat exchanger:
$$Q_1 = k \cdot \Delta T \cdot F = 749 \cdot 52.7 \cdot 0.785=11184  \ W$$

We have three such sections so approximately the heat capacity of the system is three times higher:
$$Q =33552 \ W$$

That is approximately 8\% higher than the heat capacity of the ordinary shell-and-tube heat exchanger. However, the same set of the operations was made with the conditions of laminar flow. All the initial geometry and thermal boundary conditions were the same, but the velocity of both hot and cold fluid were 0.03 m/s instead of 0.3 m/s. It changed the flow regime from turbulent to laminar and gave better results of heat capacity in the novel system. Heat capacity of the novel multi-chamber geometry system was about 30\% higher than heat capacity of the ordinary geometry system. 
Here the calculation for one section will be shown. Other sections will not be shown. Nevertheless, all the sections have their estimation in the specially created Excel spreadsheet (Figure 3). It was done to make calculations in simpler and less time-consuming way.

 \begin{figure}[H]
\centering
\includegraphics[width=0.5\textheight]{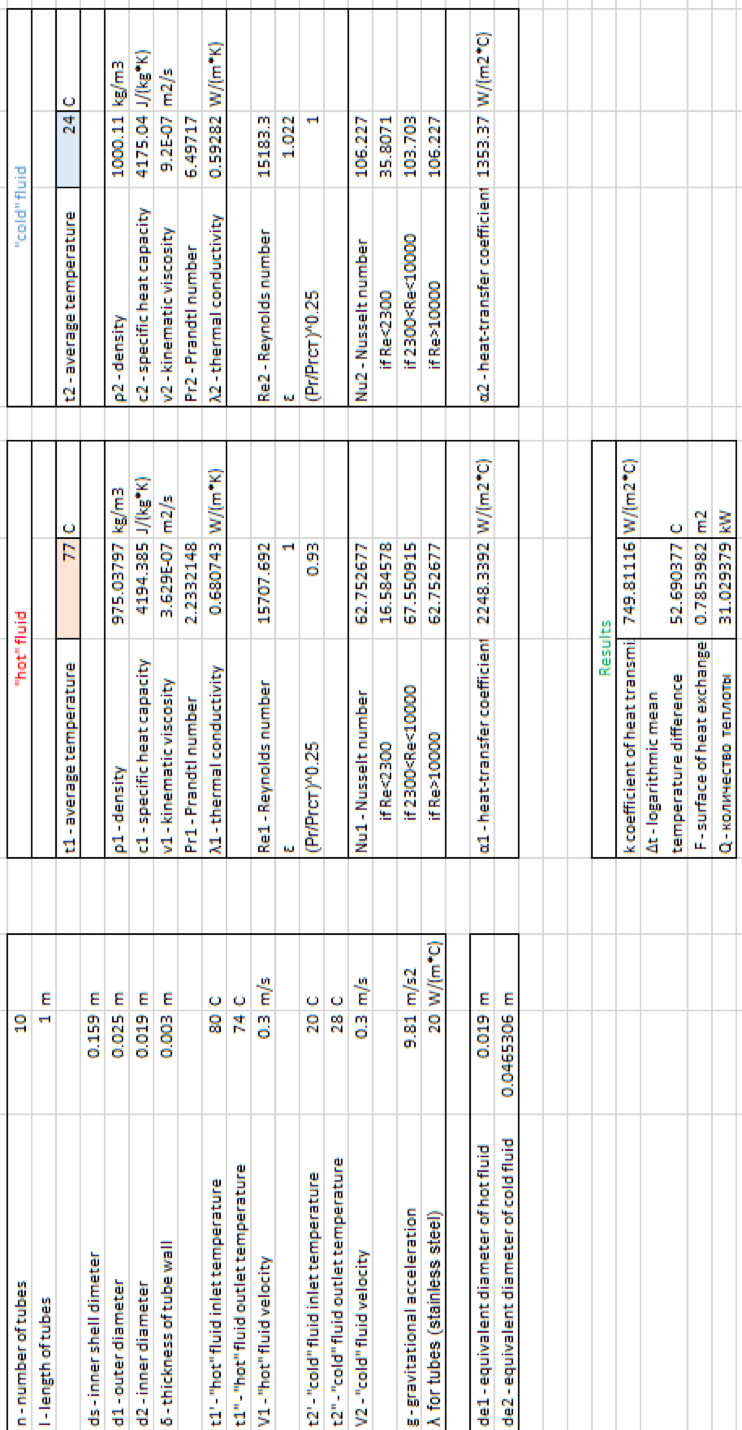}

\centering
\caption{Excel spreadsheet for the heat capacity calculation.}
\label{fig:dummy}

\end{figure}


\section{Numerical solution}

\subsection{ Geometry.} 

The geometry of two domains was prepared using SolidWorks software and can be seen on the Figure 4 -- 5. 

 \begin{figure}[H]
\centering
\includegraphics[width=0.7\textwidth]{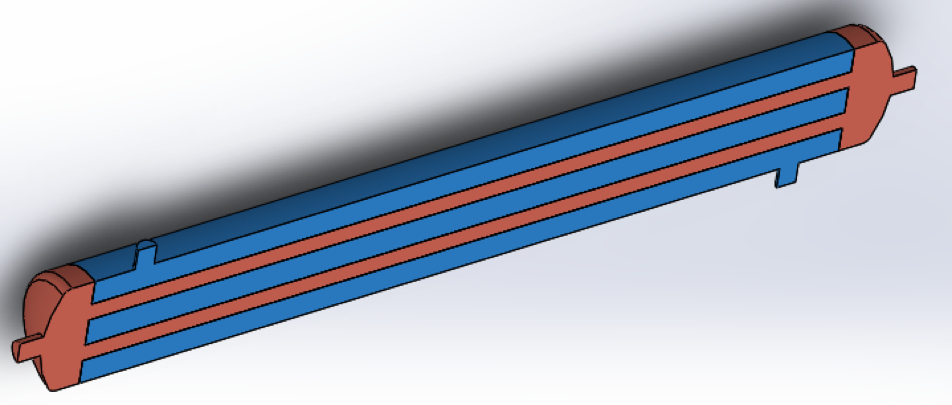}

\centering
\caption{Geometry adapted for the numerical simulation. Ordinary shell-and-tube unit.}
\label{fig:dummy}

\end{figure}

 \begin{figure}[H]
\centering
\includegraphics[width=0.7\textwidth]{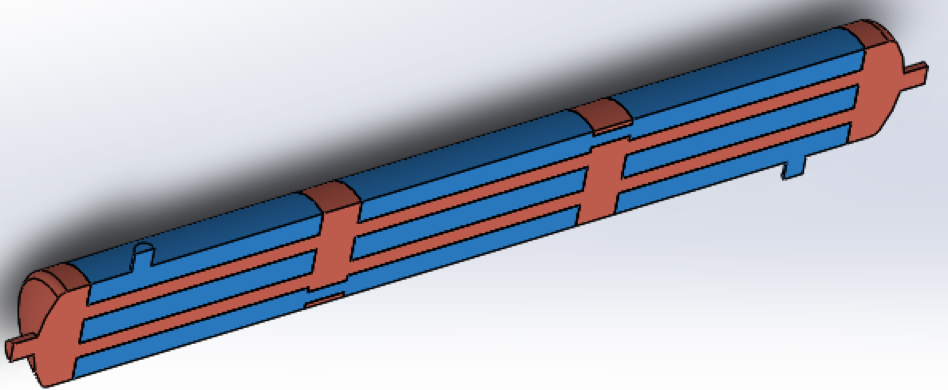}

\centering
\caption{Geometry adapted for the numerical simulation. Novel shell-and-tube unit.}
\label{fig:dummy}

\end{figure}

\medskip

\subsection{ Mesh.} 

We conduct the mesh independent study for both types of the heat exchangers and compare results for different sizes of the element of the mesh (Table 1 -- 2, Figure 6 -- 7). The method of the gradual mesh size reducing is used with the monitoring of two parameters of interest. In this case, they are temperatures of the outlet regions for hot and cold fluids [15--18]. At moment when there is no significant change between values of the temperature on two steps, it can be considered the solution is not dependent on the mesh size.

 \begin{figure}[H]

\includegraphics[width=0.3\textwidth]{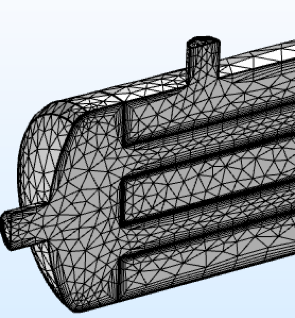}
\hfill
\includegraphics[width=0.3\textwidth]{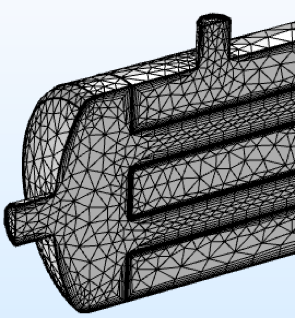}
\hfill
\includegraphics[width=0.3\textwidth]{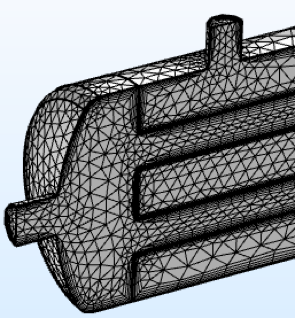}

\centering
\caption{Different mesh size for the ordinary shell-and-tube heat exchanger.}
\label{fig:dummy}

\end{figure}

\begin{table} [H]
\begin{center}
 \caption{Mesh independent study for ordinary design.}
\begin{tabular}  {| C{0.3\textwidth} | C{0.2\textwidth} | C{0.2\textwidth} | C{0.2\textwidth}|} 
 
 \hline
 \multicolumn{4}{|c|}{Type of the mesh in Comsol Multiphysics} \\
 \hline
  & Extremely coarse & Extra coarse & Coarser \\ 
 \hline
Maximum/Minimum size of the element, mm & 48.7 / 10.3 & 29.5 / 7.38 & 19.2 / 5.9 \\ 
 \hline
Number of  domain elements & 124 885 & 169 201 & 268 046 \\
 \hline
 Computational time & 2 hours 01 min & 2 hours 45 min & 14 hours 16 min \\
 \hline
 Outlet temperature of cold domain, K & 301.4 & 303.2 & 303.1 \\
 \hline
 Outlet temperature of hot domain, K & 347.8 & 346.2 & 346.0 \\ [1ex] 
 \hline
\end{tabular}
\end{center}
\end{table}

As one can see from the table, there is almost no difference between the temperatures of the second and third steps (Extra coarse and Coarser types). At the same time, the third step is more resources consuming both in terms of time and computer memory. Thereby, we can use the second step conditions for numerical simulation, considering the results of the solution have enough accuracy. The same picture with the novel shell-and tube heat exchanger (Table 2, Figure 7).

 \begin{figure}[H]

\includegraphics[width=0.3\textwidth]{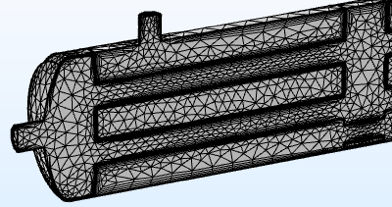}
\hfill
\includegraphics[width=0.3\textwidth]{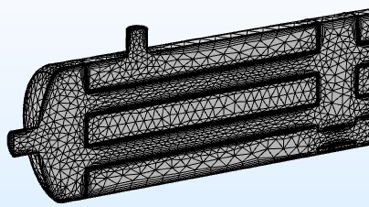}
\hfill
\includegraphics[width=0.3\textwidth]{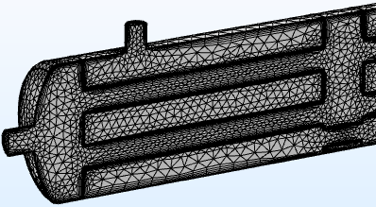}

\centering
\caption{Different mesh size for the ordinary shell-and-tube heat exchanger.}
\label{fig:dummy}

\end{figure}

\begin{table}[H]
\begin{center}
 \caption{Mesh independent study for novel design.}
\begin{tabular}  {| C{0.3\textwidth} | C{0.2\textwidth} | C{0.2\textwidth} | C{0.2\textwidth}|} 

 \hline
 \multicolumn{4}{|c|}{Type of the mesh in Comsol Multiphysics} \\
 \hline
  & Extremely coarse & Extra coarse & Coarser \\ 
 \hline
Maximum/Minimum size of the element, mm & 48.7 / 10.3 & 29.5 / 7.38 & 19.2 / 5.9 \\ 
 \hline
Number of  domain elements & 132 718 & 190 003 & 280 830 \\
 \hline
 Computational time & 2 hours 32 min & 4 hours 14 min & 17 hours 52 min \\
 \hline
 Outlet temperature of cold domain, K & 343.3 & 342.9 & 342.9 \\
 \hline
 Outlet temperature of hot domain, K & 307.2 & 308.4 & 308.3 \\ [1ex] 
 \hline
\end{tabular}
\end{center}
\end{table}

\bigskip
\subsection{ Materials and boundary conditions.} 

To have opportunity for the comparison of the performance of two types of the shell-and-tube heat exchangers, we are going to implement the same boundary conditions for both systems (Figure 8 -- 9). For numerical simulation, we need to define the inlet temperature of the hot and cold fluids and their inlet velocities. In our case, the heat transfer agents are water and water, the material of the metal part of the construction is stainless steel. Cold and hot water have 0.3 m/s velocity in the inlet region. The cold fluid has nearly room temperature and is 20$^{\circ}\mathrm{C}$ (293 K), hot water has temperature of 80$^{\circ}\mathrm{C}$ (353 K). \\
	We also implement symmetry plane. The reason for that is axisymmetric problem and reduction of the element number of mesh to get the results quicker.

 \begin{figure}[H]
\centering
\includegraphics[width=0.9\textwidth]{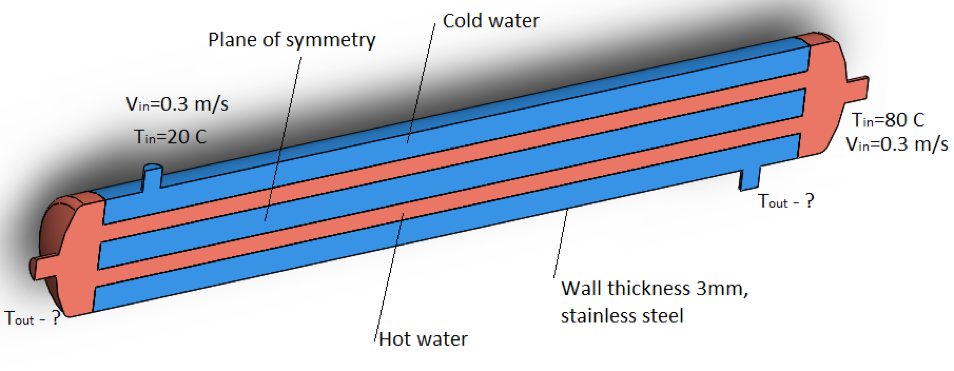}

\centering
\caption{Boundary conditions of the ordinary design of shell-and-tube heat exchanger.}
\label{fig:dummy}

\end{figure}

 \begin{figure}[H]
\centering
\includegraphics[width=0.9\textwidth]{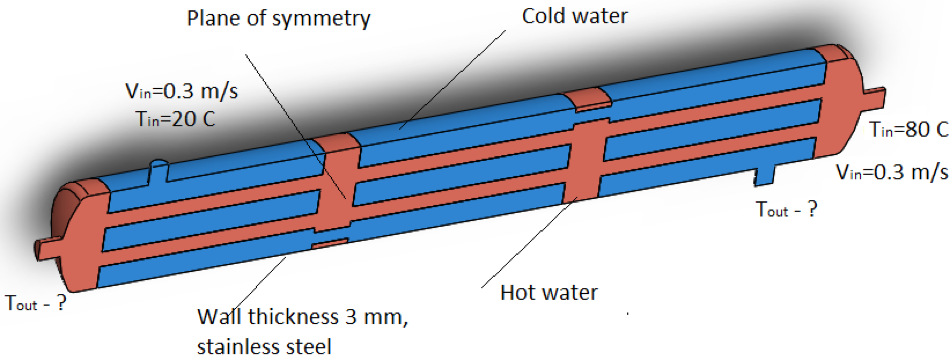}

\centering
\caption{Boundary conditions of the novel design of shell-and-tube heat exchanger.}
\label{fig:dummy}

\end{figure}

\medskip
\subsection{ Solution results analysis.} 

Problems with two different types of shell-and-tube heat exchangers were solved with the geometry, mesh and boundary conditions mentioned above. Computer was not the same as for previous sections. 
The main parameters of interest are outlet temperatures of hot and cold heat transfer agents. On the figure 10 -- 11 it can be seen the temperature distribution in the system of two fluids in thermal contact separated by metal heat transfer surface -- hot fluid domain and cold fluid domain. Figures of the temperature changing within each domain show the heating of the cold heat transfer agent and the cooling of the hot heat transfer agent.

 \begin{figure}[H]

\includegraphics[width=0.5\textwidth]{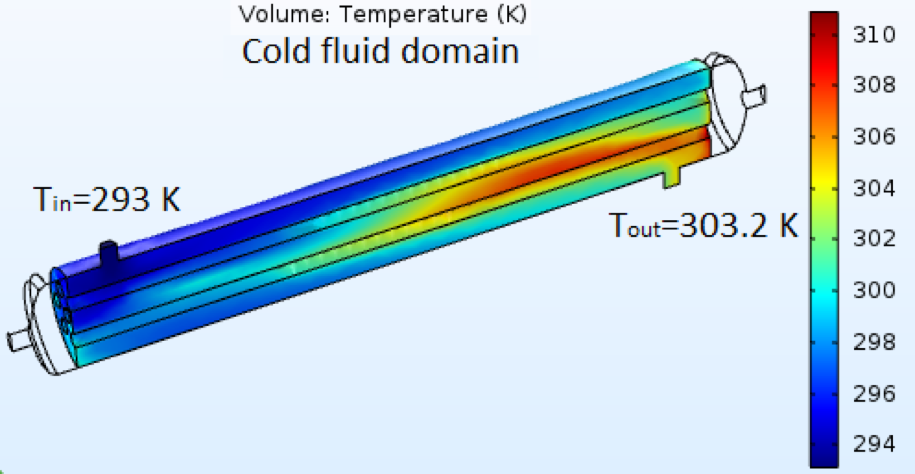}
\hfill
\includegraphics[width=0.5\textwidth]{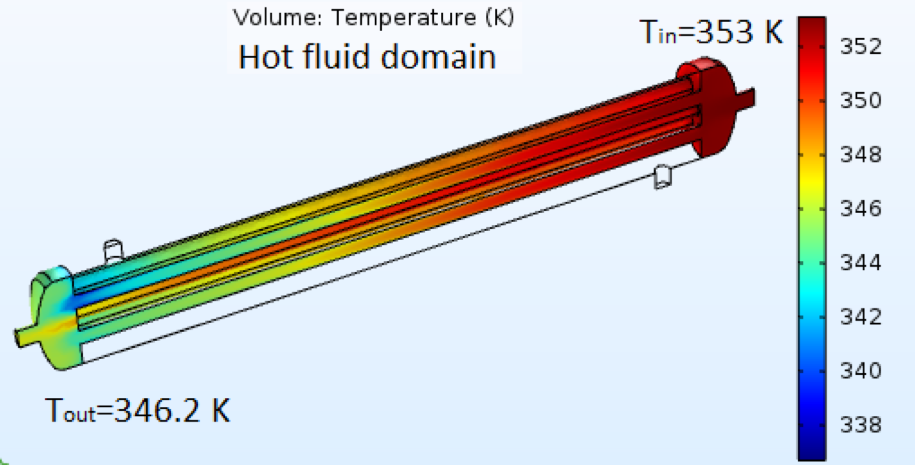}

\centering
\caption{Temperature distribution in the ordinary design of heat exchanger.}
\label{fig:dummy}

\end{figure}

 \begin{figure}[H]

\includegraphics[width=0.5\textwidth]{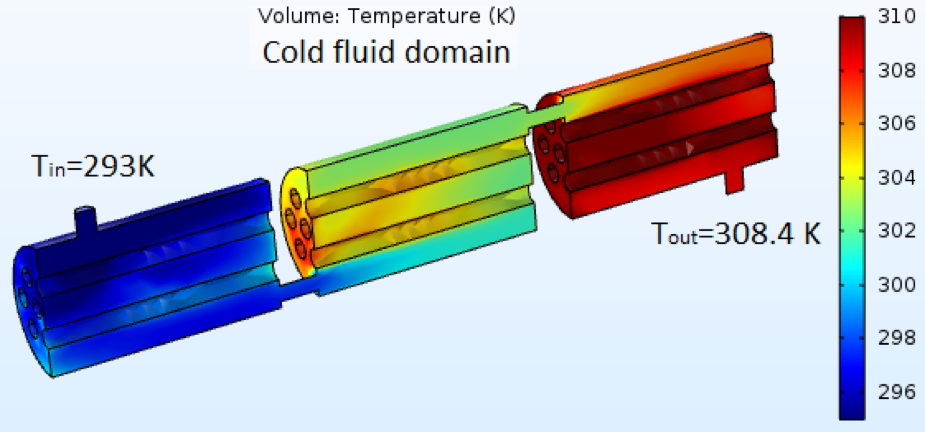}
\hfill
\includegraphics[width=0.5\textwidth]{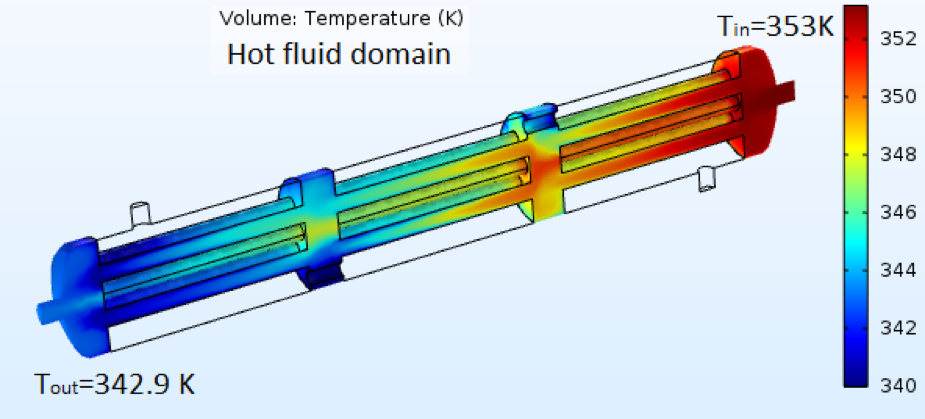}

\centering
\caption{Temperature distribution in the novel design of heat exchanger.}
\label{fig:dummy}

\end{figure}

The comparison of the results for two systems shows the better performance of the novel design of heat exchanger. The temperature difference between inlet and outlet regions of cold and hot fluids are higher in case of novel design. For the hot fluid domain, $\Delta T_1 = 353 - 346.2 = 6.8 K$ -- ordinary design; $\Delta T_2 = 353 - 342.9 = 10.1 K$ -- new design. The ratio is $\Delta T_2 / \Delta T_1 = 1.49$. The same case with the cold fluid domain, $\Delta T_1 = 303.2 - 293 = 10.2 K$, $\Delta T_2 = 308.4 - 293 = 15.4$. The ratio is $\Delta T_2 / \Delta T_1 = 1.51$. This fact allows us making conclusion about higher efficiency of the novel shell-and-tube heat exchanger. \\
However, along with the higher temperature drop novel design has the higher pressure drop (Figure 12 -- 13 -- pressure on the pictures is difference between pressure in the system and atmosphere pressure). The reason for that is complexity of the geometry. Novel heat exchanger has more regions with local resistances. In case of hot fluid domain, there is almost no difference between novel and ordinary designs. In case of cold fluid domain, the difference in pressure drop is significant. It is almost six time higher if we compare the inlet regions.

 \begin{figure}[H]

\includegraphics[width=0.5\textwidth]{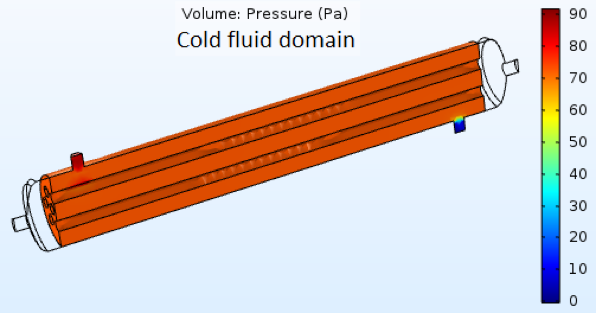}
\hfill
\includegraphics[width=0.5\textwidth]{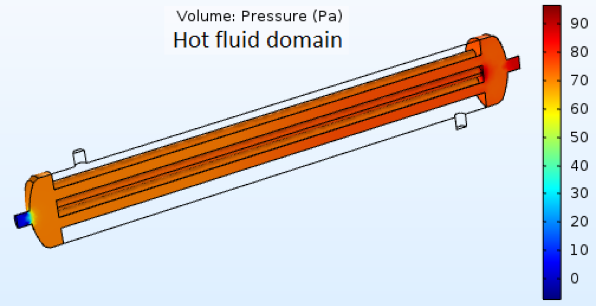}

\centering
\caption{Pressure distribution in the ordinary design of heat exchanger.}
\label{fig:dummy}

\end{figure}

 \begin{figure}[H]

\includegraphics[width=0.5\textwidth]{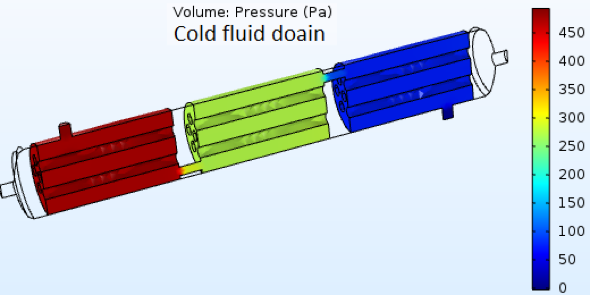}
\hfill
\includegraphics[width=0.5\textwidth]{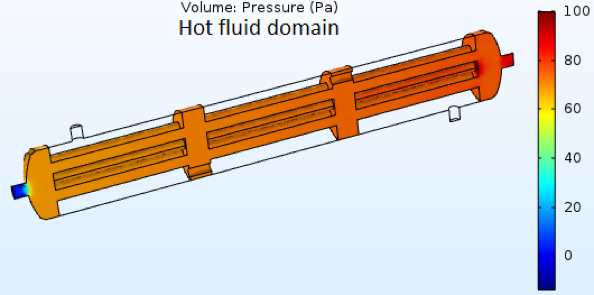}

\centering
\caption{Pressure distribution in the novel design of heat exchanger.}
\label{fig:dummy}

\end{figure}

Numerical solution shows the strengths and weaknesses of the new multi-chamber design of heat exchanger. First, it really has better performance in terms of heat capacity. If we have the system where we need to cool down the hot fluid by using cooling liquid, the novel design will give 1.5 times less temperature of interest in the outlet region of the system. At the same time, pressure drop increases almost by six times. It means that more powerful equipment for pumping of the cooling liquid is needed. \\
This obstacle needs consideration in each particular situation. For example, there are can be case when the space for the equipment installation is critical (e.g. automobile industry). The producer may not care about pumping, but the main aim is to install more heat effective exchanger unit and save the initial space. In this particular case, we do not worry about high pressure drop. However, in general case, we need to find the optimal ratio between the heat intensification and increase of the operating pressure. As it is seen from the figures above, we have the intermittent growth of the pressure in the cold fluid domain on the stage of going from one chamber into another. The reason for pressure drop is significant decrease of the cross section of the channel -- first region is shell side space, but next region is only one medium tube with several times less area of cross section. The possible solution is increase of the diameter of medium tube. Then the cross section of the shell side space will be less two. On the one hand, it leads to lower pressure step between these two regions, on the other, less cross section of the shell side leads to greater pressure from the beginning of the system. Therefore, the iteration problem needs to be solved with monitoring of the pressure drop ratio in the medium tube and chamber space and monitoring of the ratio between pressure drop and heat capacity. This process is a part of future study continuing the current one.

\section{Conclusion}
\addcontentsline{toc}{chapter}{Conclusion}

In the current study, the problem of the efficiency improvement in shell-and tube heat exchangers is investigated. Novel approach for heat transfer intensification with help of entrance hydraulic and thermal regions is described. According to hypothesis of new method, special multi-chamber geometry of the shell-and-tube heat exchanger provides better performance of the system in terms of heat capacity. This special new design was developed in order to prove the hypothesis by means of analytical, numerical and experimental solutions. \par
 The first step is analytical solution. It consists of two parts -- calculation of the heat capacity of ordinary heat exchanger and estimation of the heat capacity of novel heat exchanger. It was initial stage, but it was important to understand if the current problem and new approach of its solution really happen to be. Results of the analytical solution show 8\% higher heat capacity of the novel design in case of turbulent fluid flow, and up to 30\% higher heat capacity in case of laminar fluid flow. At this point, it is necessary to realize that the results are not one hundred percent accurate, taking into consideration all the assumptions, but they depict the characteristic behavior of the new system (efficiency improvement of the heat exchanger) and allow future speculations on this theme.\par
The second step is numerical solution. It is numerical simulation of the fluids flow inside the ordinary and novel shell-and-tube heat exchangers with the thermal contact of the fluids in the Comsol Multiphysics software environment. The problems for both types of systems were solved with the same boundary conditions in order to compare the results. Solution shows almost 1.5 times higher temperature drop of the heat transfer agents between inlet and outlet regions. It proves the hypothesis of better heat performance of the novel design. However, complex geometry of new system raises requirements for pump equipment - it should be taken into consideration during the technical-and-economic indexes estimation in each particular case.\par
The third step is experimental solution. It is set of tests on real equipment and comparison of the results -- outlet temperature measurements -- for both ordinary and novel designs. These tests will be possible only after the laboratory unit producing. Now we have almost half of the equipment ready, the rest part is on the stage of manufacturing.\par
As the conclusion it can be said that hypothesis about higher heat capacity of the multi-chamber design of shell-and-tube heat exchanger was successfully proved. Novel construction has visible advantages and can be even patented. \par


\break














\end{document}